\documentclass[10pt]{article}

\usepackage{amsmath}
\usepackage{amssymb}

\usepackage{graphicx}

\usepackage{cite}
\usepackage{hyperref}
\usepackage{color}

\usepackage{setspace}
\usepackage{soul}
\definecolor{yellow}{rgb}{0.99,0.95,0.0}

\usepackage[labelfont=bf,labelsep=period,justification=raggedright]{caption}

\makeatletter
\renewcommand{\@biblabel}[1]{\quad#1.}
\makeatother

\date{}

\begin{document}

\begin{flushleft}
{\Large
\textbf{Circadian patterns of Wikipedia editorial activity: A demographic analysis}
}
\\
Taha Yasseri$^{1\ast}$,
Robert Sumi$^{1}$,
J\'{a}nos Kert\'{e}sz$^{1,2}$
\\
\bf{1} Department of Theoretical Physics, Budapest University of Technology and Economics, Budafoki \'{u}t 8.,
H-1111 Budapest.
\\
\bf{2} BECS, School of Science, Aalto University, P.O. Box 12200, FI-00076 Espoo.
\\
$\ast$ E-mail: yasseri@phy.bme.hu
\end{flushleft}

\section*{Abstract}
Wikipedia (WP) as a collaborative, dynamical system of humans is an appropriate subject of social studies. Each
single action of the members of this society, i.e. editors, is well recorded and accessible. Using the cumulative data of 34
Wikipedias in different languages, we try to characterize and find the universalities and differences in temporal activity patterns
of editors. Based on this data, we estimate the geographical distribution of editors for each WP in the globe. Furthermore we also
clarify the differences among different groups of WPs, which originate in the variance of cultural and social
features of the communities of editors.

\section*{Introduction}\label{sec:intro} 
Relying on the data gathered by recently developed information and communication technologies (ICT), studies on social
systems has  entered into a new era, in which one is able to track and analyze the behavior of a large number of individuals and the
interaction between them in details. Among all examples, recent investigations based on cell phone records (calls \cite{karsai2011}
and text messages \cite{wu2010}) and web-based societies and media (web-pages \cite{huberman1999}, movie, news and status sharing sites e.g.,
YouTube.com \cite{szabo2010},
digg.com \cite{wu2007} and twitter.com \cite{huberman2009}) have opened very interesting insights into features of collective
and cooperative dynamics of human systems.
Wikipedia (WP) as a free, web-base encyclopedia, which is entirely written and edited by voluntaries from all around the world, has also attracted
attention of many researchers recently \cite{voss2005,wilkinson2007,ortega2008,ratkiewicz2010} (for a recent review, see \cite{park2011}). To study WP, understand  and model
its evolution\cite{voss2005}, coverage\cite{holloway2007}, conflicts or editorial wars\cite{sumi2011a,sumi2011b}, user reputation\cite{javanmardi2010}
and many other issues, we should obtain basic information about the community of its editors, i.e.,
their age, education level, nationality, individual editorial patterns, fields of interest and many other aspects.
Yet, there has been rare systematic and
unbiased studies in this direction. The main barrier here is the privacy issues, which prohibit any attempt to obtain personal data of committed editors.

There are two ways of contributing to Wikipedia. The first way is editing as an unregistered user; in this case all the edits are
recognized by the IP address of the editor, and therefore it becomes easy to locate the editor and collect some geographical
information about him/her. But most of the editors take a second way which is editing under a registered user name, which hides the
real world identity and IP address of the editors and therefore is a much more secure way of contributing. Moreover, contributions of
such serious editors are identified and unified under one single nickname, irrespective of which IP address they use to connect to the 
network and can be counted as a measure of
maturity in the promotion processes. Cohen has extracted geographical data from
IP addresses of unregistered editors of English WP, integrated them over time and concluded that about $80\%$ of edits on English WP are originated
from few English speaking countries with high Internet penetration rate, i.e., $60\%$ from the USA, $12\%$ from UK,
$7\%$ from Canada and $5\%$ from Australia \cite{cohen2010}.
However, contributions from unregistered editors are
limited to less than 10 percent for many WPs (see Table~\ref{tab:stat}). Moreover the rather small sample of unregistered users,
is not representing the features of average users, as will be discussed later. Therefore, indirect methods to
locate editors or to obtain
any kind of information about the community is highly desirable. 
One of our aims is to show that using the temporal patterns of WP users, conclusions about the geographical distribution of (registered) editors can be drawn.

Recently much effort has been devoted to describe and understand the extreme temporal inhomogeneity of human activities, represented by
the burstiness of activities and the fat-tailed distribution of time intervals between
events\cite{barabasi2005}. While the circadion and other periodic characteristics of temporal patterns of human activities cannot account for the whole richness of bursty behavior\cite{jo2011}, they remain important for understanding the entire dynamics of the systems. These regularities
are induced by circadian and seasonal cycles of the nature \cite{panda2002} on one hand and by cultural aspects on the other one. Consequently,
studies on diurnal patterns of the Internet traffic have brought interesting information about individual habits of the Internet usage
in different societies \cite{spennemann2006,spennemann2007}.
In this paper our focus is on such cyclic behavior, while investigations on other aspects of temporal inhomogeneities like short time bursty behavior and
inter-event interval distributions are reported elsewhere \cite{yasseri-po}.

West et.~al. have tried to make use of diurnal characteristics of edits to detect vandalism and destructive edits \cite{west2010}. Their study was
again restricted to tracking positive and negative edits from
unregistered editors, for which they found that most of the "offending edits" are
committed during the working hours and working days compared to after-darks and
weekends.
In the admin-ship of Wikipedia it is also becoming fashionable to use the personal temporal fingerprint of editors as a side-tool to
detect and prevent sock-puppetry, \footnote{WP editors are generally expected to edit using only one account. Sock puppetry is the use of
multiple accounts to deceive other editors, disrupt discussions, distort consensus,
avoid sanctions, etc., which is according to WP rules forbidden.}
although this could only be done with high respect to the privacy policies of Wikimedia
Foundation.\footnote{\url{http://wikimediafoundation.org/wiki/Privacy_policy}}

In this work, we first try to characterize the circadian pattern of edits on Wikipedia,
by analyzing massive data of 34
WPs, then we introduce a novel method to locate and find geographical distribution of
the editors of large international WPs, e.g., English, simple English, Spanish, etc.
Furthermore, we analyze the temporal behavior of editors on longer time scales, i.e. weekly patterns and report on significant differences
between various societies.

\section*{Methods}\label{sec:data}
This work is carried out on 34 WPs selected from the largest ones in
respect to the number of articles, i.e., those ones, which have more than 100,000 articles.\footnote{Two Wikipedias of Volap\"{u}k and
Waray-Waray are excluded from the list due to their small number of speakers and Wikipedians and considering that many of articles are robotically generated.
The simple English Wikipedia is also included in the list, despite it contains only around 70,000 articles.}
Among the sample, number of total edits and editors vary between 3 M to 455 M  and 46 k  to 14 M,  respectively.
In Table~\ref{tab:stat} some statistics about the WPs under the investigation are reported.

We considered every single edit performed on each WP and having the timestamps assigned to edits,
calculated the overall activity of users for the time of day and day of the week. To see the universality
of circadian activity patterns among editors of all different languages, we assumed a local time offset for
each language. Clearly there are some languages which are not spoken only in one country or one time zone, e.g.,
Spanish, Arabic, etc, whereas some others are very localized in a specific time zone, e.g., Italian, Hungarian, etc.
For the first sort of languages, we initially considered the time offset of the most known origin of the language.
For the special cases of the English and simple English Wikipedias, initially we considered an offset corresponding to
USA Central Time. In the ninth column of Table~\ref{tab:stat}, the assigned time offset to each language is reported.
Note that, due to lack of information such as IP addresses of users, this initial assumptions for the origin of edits and
corresponding time offset can not be any improved at this step.
It is one of our goals to implement a method, based on the average behavior of WP editors, which is able to determine the percentage of the contributions coming from different geographic units. This method will be described in the next section in sequence with the empirical observations.


\section*{Results}\label{sec:results}
\subsection*{Circadian patterns}\label{subsec:circadian}
We calculated the normalized number of edits for each of the 34 WPs with the consecutive time windows of one hour for
 the 24 hours of the days. The rational activity level of each time window is calculated by dividing the number of edits within the time window by the
total number of edits.

This way the circadian activity patterns are created as
depicted in Figure~\ref{fig:daily} (a). Most WPs show a universal pattern;
A minimum of activity at around 6 A.M., followed by a rapid increase up to noon. The activity shows a slight increase
until around 9 P.M., where it start to decrease during night.
Qualitatively similar shapes are observed for other kind of human activities, e.g. cell phone callings and textings\cite{jo2011}, and
the Internet instant messaging \cite{pozdnoukhov2010}.

\subsubsection*{Deviations}\label{subsubsec:dev}
\begin{figure}[!ht]
\begin{center}
\includegraphics[width=0.95\textwidth]{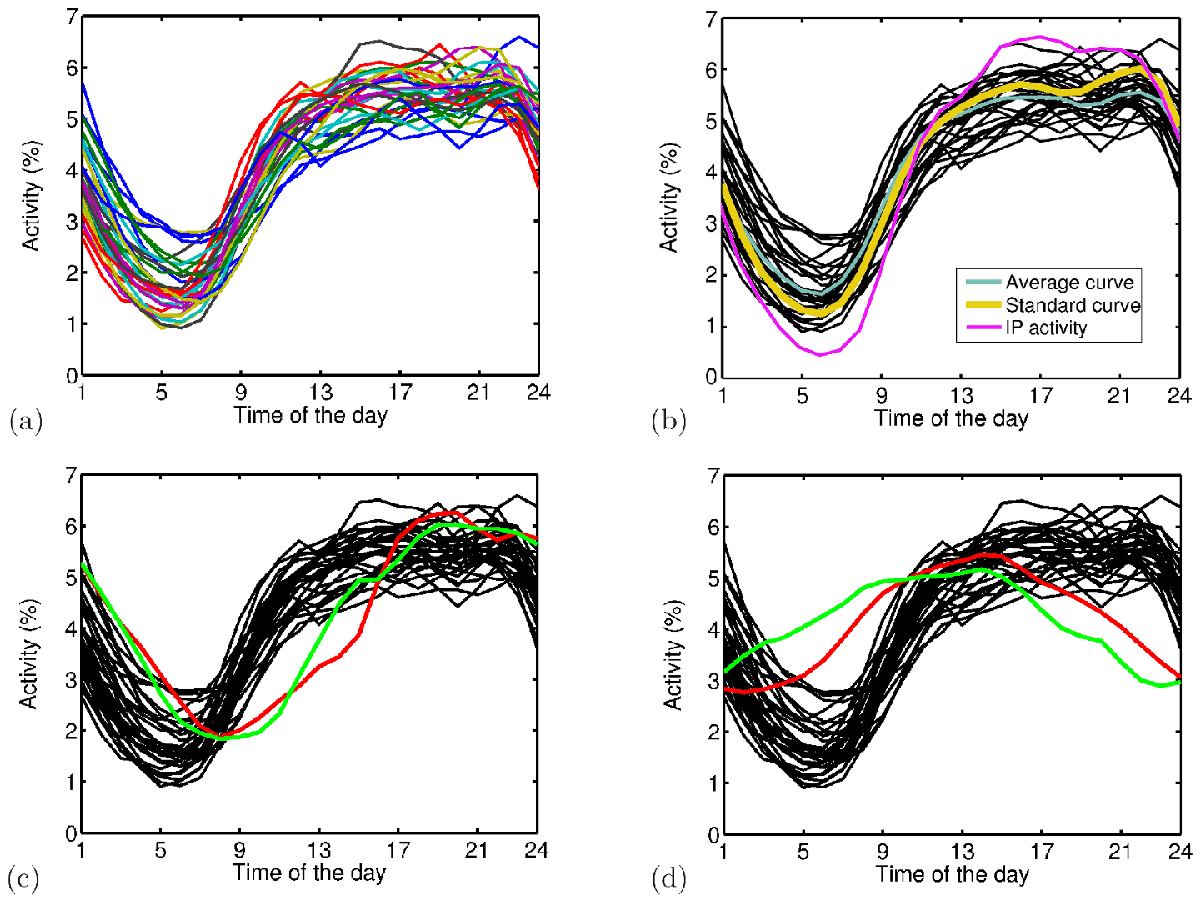} 
\end{center}
  \caption{{\bf Normalized activity of editors} for (a) all WPs listed in Table~\ref{tab:stat}
  excluding English, simple English, Spanish and Portuguese, (b) the average curve extracted 
from curves in (a) and standard curve extracted
from 10 most localized WPs along with the activity curve of unregistered users, whose IP-addresses are known and
therefore one is able to locate them and obtain the local time zone precisely (c) activity pattern  of Spanish
  (red) and Portuguese (green) WPs, and (d) activity pattern of English (red)
  and simple English (green) WPs.} \label{fig:daily}
\end{figure}

Among all 34 investigated WPs, there are four, which significantly deviate from all the others in respect to the circadian patterns.
in Figure.~\ref{fig:daily} (c) and (d) diurnal activity for these four outliers, Spanish, Portuguese,
English and simple English WPs are shown. In the case of Spanish and Portuguese, the main difference
to the rest WPs, is the slight shift to the right (later times). Having in mind that Spain and Portugal
both use local times which have a larger offset compared to the countries with the same longitude,
this comes as no surprise. Beside that, the rather large number of speakers from Latin America not only
is in favor of this shift, but also flatten the overall amplitude of the diurnal pattern (this
will be discussed later in more details). And finally the cultural features of those two countries
might contribute to this observation.


In the case of the English and simple English WPs, for simplicity, we assumed the reference being UTC-6 (which corresponds to the Central Time Zone of the US). Naturally 
the deviations from the universal pattern are very strong, indicating the complex origin of the English WP. Later we will come back to this point.

\begin{figure}[!ht]
\begin{center}
\includegraphics[width=0.95\textwidth]{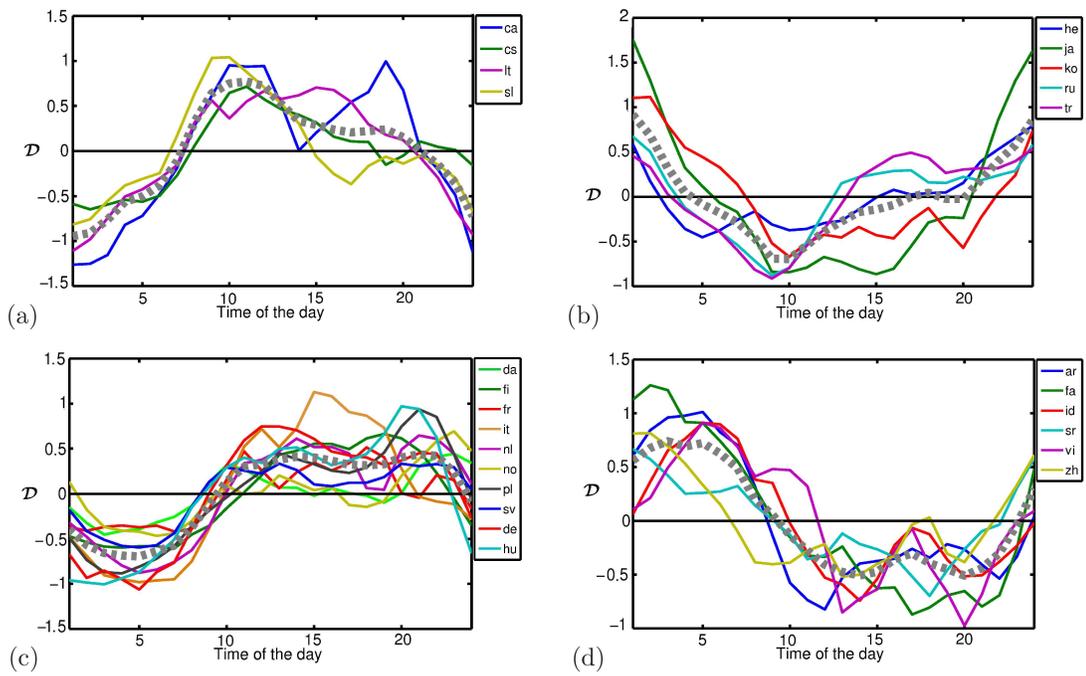} 
\end{center}
  \caption{{\bf Deviation of activity patterns from the average curve,} leading to 4 different categories of WPs. The gray dotted line is the average deviation
of each category. The sleep depth (for the definition, see the text) of for categories a-d are 4.5$\pm$0.2, 4.3$\pm$0.2, 4.9$\pm$0.1 and 2.8$\pm$0.2 respectively.} \label{fig:dev}
\end{figure}

To better illustrate the deviation from an average circadian pattern, we calculated the weighted average of curves in Figure.~\ref{fig:daily} (a).
Each WPs pattern is weighted by its total number of edits. The average curve is depicted in Figure.~\ref{fig:daily} (b). Now we can calculate
the difference from this average pattern for each WP, $\mathcal{D}(t)$ at different times of the day $t$. According to the shape of $\mathcal{D}(t)$ and by
maximizing the cross-correlation coefficient, almost all WPs
could be categorized in 4 categories as in Figure.~\ref{fig:dev}. Two of these categories, Figure.~\ref{fig:dev} (a) and (c)  consist of WPs which have
less activity during nights compared to the average pattern.

These WPs are all in such European languages, which are spoken in single, localized regions and therefore the
minimum of activity of their editors is deeper than others. In Figure.~\ref{fig:dev} (b), a category consisting of
Asian languages is shown. These WPs are more active during nights and less active during working hours compared to the average.
In the last category, shown in
Figure.~\ref{fig:dev} (d), a higher activity during night and a lower activity during working hours is a clear sign of a extended distribution of contributors
from different time zones. Arabic, Persian, Chinese are from this category in addition to Spanish, Portuguese, English and Simple English (not shown).

The other way to look at the locality of the languages is to quantify the {\it sleep depth}.
Sleep depth is defined as the
difference between the maximum and the minimum of the activity of each language users and
might be assumed as a measure of the locality of the global distribution of the
editors of the corresponding language. In the last column of Table~\ref{tab:stat}, the calculated depth values are
reported. These values vary from 2.3 for simple English to 5.6 for Italian. Among those WPs with small sleep
depth are Arabic, Indonesian, Persian and English. The average sleep depth for the category of Fig.~\ref{fig:dev}(d) is $2.8$ with standard deviation of $0.4$.

Among languages with the large sleep depth are Italian, Hungarian, Polish, Catalan, and Dutch. These are all languages which are
mostly spoken in a narrow area of the world and therefore are very localized in time zones.
The average sleep depth for the category of Fig.~\ref{fig:dev}(c) is $4.9$ with standard deviation of $0.4$.
It is also to mention, that although Spanish and Portuguese are both widely spoken in different areas and different time zones, but
the sleep depth of both lay in the middle range (4.4 and 4.2 respectively). For a more precise interpretation we try to estimate the
share of editors from different areas to each WP in the next section.

\subsubsection*{Geographical distribution of editors}\label{subsubsec:spat}

As mentioned above, due to privacy policy issues, there is no access to the locating information of registered editors, such IP addresses. However there
are studies only considering contributions by unregistered users which give a very rough image of the real distribution
 of editors in the globe\cite{cohen2010}.
We aim at a better method by decomposing the overall activity pattern of each WP to basic elements, which are assumed to be representative for
contributions purely originated from a certain time zone. For this purpose, we averaged over activity patterns for the 10 WP with the deepest sleep to obtain
a smooth curve, which has the features of collective activity of users in synchrony (hereafter called Standard curve $\mathcal{S}(t)$).
In the next step, we assume that the activity pattern of a WP, $\mathcal{A}(t)$ with wide spatial distribution of editors can be
simulated by superpositions of $N$ standard
curves with different time shifts $\tau_i$ and different weights $w_i$ for $i=1$ to $N$,
\begin{equation}
 \mathcal{A}(t) = \sum\limits_{i=1}^N w_i \mathcal{S}(t-\Delta \tau_i)
\end{equation}
where $\Delta \tau_i$ is the difference between $\tau_i$ and the assumed time offset of the language (see Table.~\ref{tab:stat}).

\begin{figure}[!ht]
\begin{center}
\includegraphics[width=0.7\textwidth]{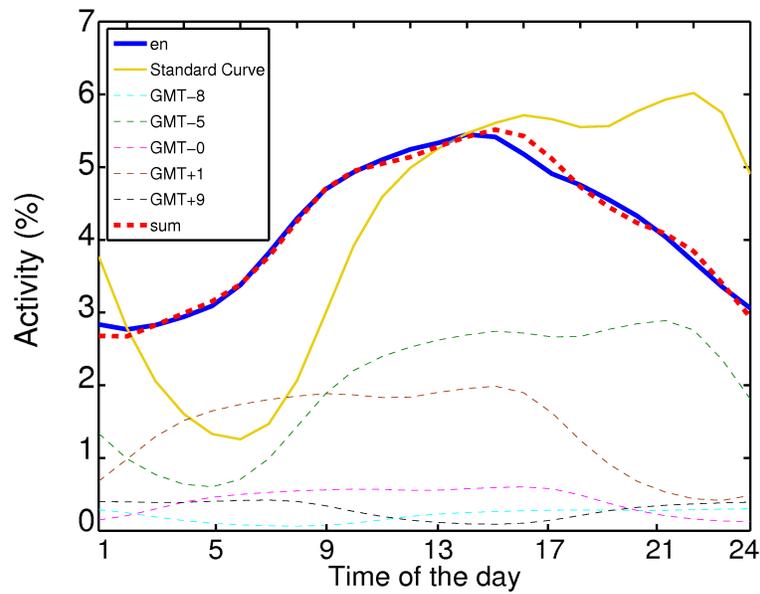}
\end{center}
  \caption{{\bf Decomposition of activity pattern of English WP into 5 shifted standard curves with different weights.} the blue line is the empirical data,
the yellow curve is the {\it standard curve} (see the text for the definition), the thin dotted lines are shifted and weighted standard curves, and the red dotted
line is the linear superposition of them which models the empirical data properly.}\label{fig:dc}
\end{figure}

\begin{figure}[!ht]
\begin{center}
\includegraphics[width=0.95\textwidth]{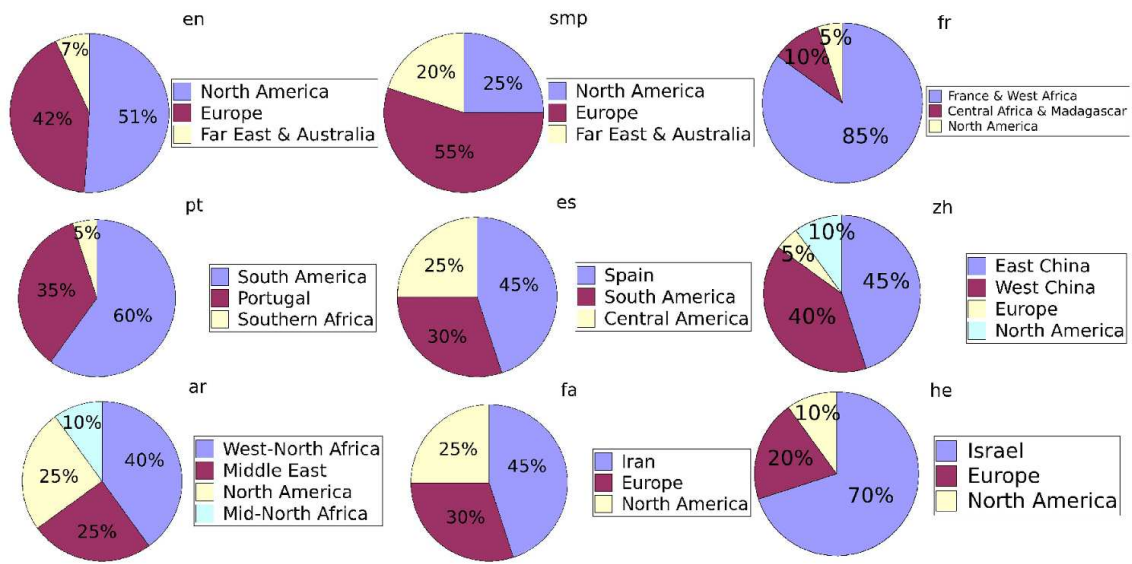}
\end{center}
  \caption{{\bf Estimation of users contribution from different regions.} By precisely combining the outputs of the optimization process, described in
the text, and demographic data of each language, the share of each region to each WP is estimated. For the sake of accuracy in reporting the results,
 in some cases the contributions of regions closely located, are unified.} \label{fig:estimate}
\end{figure}

In general, one could minimize the error of the simulated activity pattern for each WP for $N=24$ different offsets and find the optimal weighting.
Clearly, weights are proportional to the volume of contributions from each time zone. Following this outline we did the optimization, but in a more
supervised manner. We restricted $N$ to the number of different time zones, which are relevant candidates for being an origin of contribution, e.g., we excluded
time zones of nonliving areas of the earth. Furthermore, to reduce the complexity of calculations and also avoid multiple solutions, we reduced $N$ to
the number of areas, which have considerable number of speakers of the language. In many cases, by superposition of $N$ between 3 to 6 standard curves,
we could fit the empirical data with a high value of correlation coefficient between the simulated and imperial data sets (see Figure~\ref{fig:dc})
, whereas taking larger $N$s does not decrease the error
and it only leads to more zero $w_i$'s. Finally, by a proper combination of demographic information and optimization techniques, we
estimated the share of different regions to 9 different WPs. These estimations are summarized in Figure~\ref{fig:estimate}.
Though in some cases the error function is rather flat around its minimum, leading to relative large tolerance in calculated weights, existence of separated multiple minimums is
prohibited by applying the demographic restrictions.


\subsection*{Weekly patterns}\label{subsec:weekly}

\begin{figure}[!ht]
\begin{center}
\includegraphics[width=0.95\textwidth]{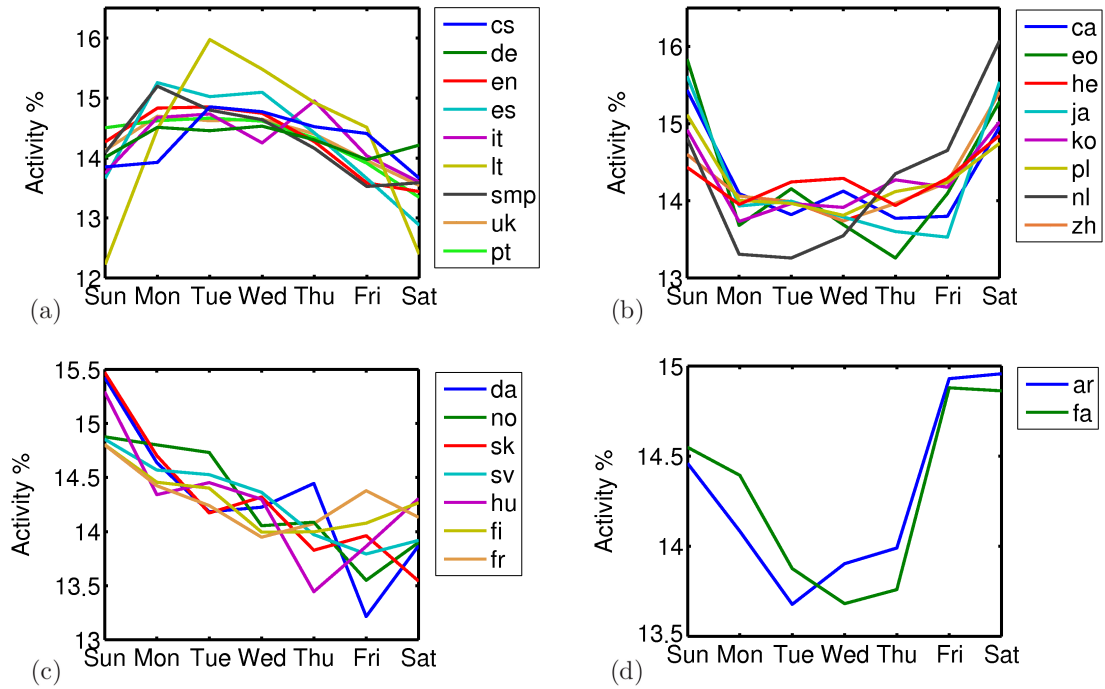}
\end{center}
\caption{
{\bf Activity of editors on different days of the week, categorized in 4 subcategories.} There are two major categories of weekend, (b)-(d) and weekdays (a)
activity. Please note that WPs in (d) are languages spoken mostly in Muslim countries, which have either Thursday–Friday weekend (Saudi Arabia,Oman and
Yemen), or Friday–Saturday weekend (Algeria, Bahrain, Egypt, Iraq, Jordan, Kuwait, Libya, Qatar, Sudan,Syria, United Arab Emirates), or
, Saturday–Sunday weekend (Morocco, Tunisia). In Iran and Afghanistan, Persian spoken countries, only Friday is considered as weekend.} \label{fig:weekly}
\end{figure}

We also considered the activity of editors during weeks and its dependence on the day of the week. These results are shown in
Figure~\ref{fig:weekly}. According to the weekly pattern of activity, we could categorize 28 out of 34 WPs into 4 different
categories which belong to two main categories of "working days" and "weekend" activity. In the upper-left panel of
Figure~\ref{fig:weekly}, those WPs are shown, which have highest activity of editors during the working days of the week.
Among them, are English, simple English, German, Spanish, Portuguese and Italian. In the rest of WPs, a big part of edits
are done during weekends. In the class of Polish, Dutch, Korean and Japanese WPs (upper-right panel of Figure~\ref{fig:weekly})
equal activities are shown on Saturdays and Sundays, whereas in the class of Danish, Swedish, Norwegian and Finnish WPs, editors
have very low activity on Saturdays. The last class of the "weekend" WPs, consists of Arabic and Persian WPs, in which Fridays
are also active days in addition to Saturdays and Sundays. The latter is no surprise, considering that Friday is a public
holiday in all of the original countries along with Saturday in most of them.

\section*{Discussion}\label{sec:discussion}

The novel approach to the collective characteristics of community of editors of WPs, described above, enables us, for the first time, to shed light
on less studied aspects of Wikipedia. Based on the reported results, many basic questions and concerns about the whole projects of Wikimedia
can be investigated. Knowing the spatial distribution of the editors of a certain WP would be reliable basis for explaining specific
biases in WP articles, heterogeneous topical coverage and origins of conflicts and editorial wars to some good extent. In addition to that, these results
arise new questions and puzzles as well.
Considering the large population of
English speakers in North America compared to Europe, and the fact that the Internet
is most developed in North America, the estimation of around only half share for north America to English WP is a puzzle, which definitely
needs further multidisciplinary studies. In the case of Simple English WP, the European share is even larger, which is not surprising, together with the fact that the share of Far East increased, since this WP is
meant to be of use by non-native speakers (though, not necessarily written by them). Note that previous results of
\cite{cohen2010} and \cite{west2010} are partially supported by the results reported here.
For instance, a share of less than $10\%$
for Australian editors in
English WP is in both articles reported. Unfortunately, there is no explicit focus on the contributions from European countries in the mentioned works, and
it seems the large amount of efforts by European editors was overlooked. However we have repeated the measurements on 
IP addresses of unregistered users more generally for different WPs by following every single edit from this type to locate
the editor. Firstly, we constructed the ``precise`` activity pattern of unregistered users, as shown in 
Fig.~\ref{fig:daily}(b). The activity pattern of unregistered users has clearly deeper minimum at night and higher maximum
during working hours, compare to most of the other curves. Unregistered users contribute to WP occasionally and mostly only
with few edits from the same IP address. To be actively editing even at nights, one must be extremely committed to WP, 
therefore
the deep sleep of the activity curve of unregistered users comes as no surprise. We believe the sample of unregistered users
is more representing the activity of WP readers who edit rarely as they notice needs to tiny modifications here and there
while reading the articles than committed users who basically write the main body of the articles. 
The percentage of contributions by 
unregistered users is measured and reported in Table~\ref{tab:stat} for all 34 WPs. This value varies between 4 for Slovenian 
and 37 for Japanese WP. We compared the results for 
geographical distribution of editors obtained
by locating them with the IP addresses to the previous results described above and observed that, both methods mainly 
give similar results for the WPs with rather larger share of unregistered users, whereas they deviate for WPs with
 small share of unregistered users. Finally, one should consider the fact that the committed users, sometimes edit without 
using their registered user name to vandalize or edit specific controversial articles without leaving any trace which 
may cause troubles
for the original user name. In such cases, most of the time an ''open proxy`` (with an arbitrary IP address) 
is used to hide the real IP address of the editor.
This makes the analysis based on IP addresses even looser.

Another interesting part of the results is on Persian WP. Although more than $70 \%$ of native Persian speakers
live in Iran and the rest in closely neighboring countries, but the corresponding WP appears in the top list of 
WPs with small sleep depth. In addition, the estimated share for edits from Iran is only
about $45 \%$. This could be due to the following facts. 1) Strong restrictions on the Internet usage have been applied by Iranian government during years
as a consequence of socio-political issues\footnote{Wikipedia is one of the few remaining 
unbanned Web.2.0 web sites currently in Iran.}, 
which makes it difficult to contribute to WP using Iranian based ISPs. 2) Iran has a high rate of immigration of
students and scholars. That has led to formation of large intellectual communities out of Iran, which might be
responsible for considerable amount of edits in Persian WP.

Low level of contribution to the French WP by North American editors and to the Arabic WP by Egyptian editors, could have roots
in the differences between the spoken dialects and the standard languages.
Though both languages (French and Arabic) are among the official languages in the mentioned regions, it seems that the divergence
between dialects play an important role to suppress contributing to WP. It should be mentioned, that there is a separate WP in the local dialect of Egypt,
({\it Egyptian Arabic Wikipedia}) and there has been an unsuccessful effort to launch the {\it Canadian French Wikipedia} recently. Therefore we think that
the estimations for contributions could be of interest for the WP community too and elaborate the process of decision making for a new WP in a local dialect.

Clearly, the presented method also has its limitations. For instance, accessing to information about the distribution of editors in different
longitudes is impossible by only considering the time stamps. Moreover the resolution of the regional estimations are not very high. Because of many
factors, e.g. applying summer time in many countries the method can not claim at a resolution higher than a one hour stripe. For example,
in the case of English WP, the supervised optimization results in a ratio of 3 for the weight of GMT+1 over GMT+0, corresponding to Central Europe and Western Europe times. But
because of the mentioned reasons, distinction between the share of the very close time zones is not justifiable. Moreover, in some cases the error of the
simulated activity pattern is not very sensitive to changes in weights of spatially closed offsets. However, all the 
results presented above are precise
up to the last significant digit.

Putting beside the deviations from the average of daily activities and the weekly activity for all WPs, one is able to make very clear conclusions. For example, the daily
pattern of Asian languages (e.g., Japanese, Chinese and Korean) show higher activity during evenings and nights along with high level of activity at weekends.
This can be related partly to the lengths of working hours in corresponding countries. This general image, which holds partially for Turkey and Russia
and Israel too,
could
be in close relation with the high average working hours per day (more than 40 hours in all the mentioned cases\footnote{According to the dataset of
{\it The Organization for Economic Co-operation and Development}: \url{http://stats.oecd.org}}) in those countries. Furthermore, among European countries,
we also see the same tendency; in the countries with rather larger working times, edits are mostly done in later times in evenings.

It is to mention that same analysis have been done for the seasonal patterns to extract effects of changes in daylight timing, but the large fluctuations
in average behavior, makes it very difficult to conclude relevant results. The only significant large scale seasonal pattern is the reduction of
activity with approaching to the new year holidays for many WPs.

In conclusion, based on a dataset of time stamped edits on different Wikipedias, we studied the diurnal and weekly patterns of activity of
editors. We could see a universal circadian pattern for all WPs, which has its minimum at dawn and maximum at late afternoon
and early evening. According to this investigation, we also argued that using a weighted mixture of contributions from different time zones and an optimization procedure, we can estimate the different contributions to a WP.
In particular, we observe that
a considerably large part of edits on English and simple English
WPs are originated from Europe and the share of North America was below expectations. The same type of analysis was also performed
for other WPs in different languages.
In contrast to diurnal pattern, which is universal to a great extent, weekly activity patterns of WPs show remarkable differences. We could, however, identify two main categories, namely "weekends" and "working days" active WPs. Further studies are needed to explain these observations in detail and relate
them to cultural and social differences.

\section*{Acknowledgments}
The project ICTeCollective acknowledges the financial support of the Future and Emerging Technologies (FET) program within the Seventh Framework Program for Research of the European Commission, under FET-Open grant number: 238597. JK and TY thanks the FiDiPro program of TEKES for partial support. We also thank Wikimedia Deutschland e.V. for providing us with the data through the Wikimedia toolserver platform.

\bibliography{circadian}

\section*{Figure Legends}

\section*{Tables}
\begin{table}[!ht]
\caption{
\bf{Wikipedias Statistics}}
\begin{tabular}{|c|c|c|c|c|c|c|c|c|c|c|}
\hline WP	&Language	&M.~Country	&Speakers	&Articles	&Edits	&Users	&Active	&IP\%	&Offset	&S.~D.\\
\hline ar	&Arabic	&$^*$	&255	&146	&8	&368	&2433	&12	&+2$^\ddag$	&2.5\\
\hline bg	&Bulgarian	&Bulgaria	&9	&115	&4	&89	&848	&14	&2	&4.4\\
\hline ca	&Catalan	&Andorra	&7	&337	&7	&85	&1749	&6	&0	&5.2\\
\hline cs	&Czech	&CzechRep.	&11	&192	&6	&146	&2329	&9	&1	&4.3\\
\hline da	&Danish	&Denmark	&5	&147	&5	&128	&1364	&10	&1	&4.3\\
\hline de	&German	&Germany	&128	&1214	&91	&1205	&24519	&16	&1	&5\\
\hline en	&English	&-$^*$	&600	&3609	&455	&14340	&151549	&10	&-6$^+$	&2.7\\
\hline eo	&Esperanto	&-$^\dag$	&2	&143	&3	&49	&466	&6	&+1$^\blacktriangle$	&3.5\\
\hline es	&Spanish	&-$^*$	&460	&748	&48	&1789	&15647	&26	&+1$^\vartriangle$	&4.4\\
\hline fa	&Persian	&Iran	&110	&124	&6	&217	&1890	&5	&3.5	&2.7\\
\hline fi	&Finnish	&Finland	&5	&266	&10	&175	&2060	&16	&2	&4.8\\
\hline fr	&French	&France	&172	&1088	&68	&1037	&16546	&14	&1	&4.5\\
\hline he	&Hebrew	&Israel	&5	&116	&11	&140	&2020	&31	&2	&4.5\\
\hline hu	&Hungarian	&Hungary	&13	&187	&10	&167	&2055	&6	&1	&5.3\\
\hline id	&Indonesian	&Indonesia	&160	&159	&5	&247	&1881	&9	&8	&2.6\\
\hline it	&Italian	&Italy	&62	&790	&43	&620	&8279	&18	&1	&5.6\\
\hline ja	&Japanese	&Japan	&126	&743	&37	&510	&10571	&37	&9	&4.8\\
\hline ko	&Korean	&Korea	&67	&159	&7	&147	&1916	&14	&9	&3.4\\
\hline lt	&Lithuanian	&Lithuania	&3	&131	&3	&46	&497	&7	&2	&4.6\\
\hline nl	&Dutch	&Netherlands	&20	&681	&25	&381	&5125	&10	&1	&5.1\\
\hline no	&Norwegian	&Norway	&4	&297	&9	&194	&2413	&8	&1	&4.7\\
\hline pl	&Polish	&Poland	&44	&793	&27	&425	&5403	&14	&1	&5.2\\
\hline pt	&Portuguese	&-$^*$	&230	&680	&25	&852	&5770	&20	&0$^\S$	&4.2\\
\hline ro	&Romanian	&Romania	&27	&158	&5	&181	&1255	&7	&2	&3.6\\
\hline ru	&Russian	&Russia	&277	&699	&35	&652	&12841	&16	&3	&4.3\\
\hline simple	&sim.English	&-$^*$	&-	&69	&2	&176	&746	&13	&-6$^+$	&2.3\\
\hline sk	&Slovak	&Slovakia	&5	&122	&3	&58	&612	&7	&1	&3.7\\
\hline sl	&Slovenian	&Slovenia	&1	&109	&2	&78	&579	&4	&1	&3.8\\
\hline sr	&Serbian	&Serbia	&11	&141	&4	&81	&672	&5	&1	&3.5\\
\hline sv	&Swedish	&Sweden	&9	&392	&14	&221	&3467	&14	&1	&4.5\\
\hline tr	&Turkish	&Turkey	&75	&158	&9	&337	&2499	&24	&2	&4.5\\
\hline uk	&Ukrainian	&Ukraine	&37	&274	&6	&99	&1929	&5	&2	&4\\
\hline vi	&Vietnamese	&Vietnam	&86	&201	&4	&223	&1156	&8	&7	&2.7\\
\hline zh	&Chinese	&China	&1300	&351	&16	&984	&5696	&13	&8	&3.7\\
\hline
\end{tabular}
{\small
$^*$For the languages which are widely spoken in the world, the origin country is not well-defined.
$^\dag$Esperanto has never been an official language of any country.
$^\ddag$Egypt (the most populated Arab country) time zone.
$^+$USA Central standard time zone.
$^\blacktriangle$Central European time zone.
$^\vartriangle$Spain time zone.
$^\S$Portugal time zone.}
\begin{flushleft}Statistics about WPs under investigation. Name of the WP, language, the most populated country, in which the language is spoken,
and total number of speakers in the world (millions) are reported in columns 1 to 4, followed by number of articles (thousands)  in the WP,
number of edits (millions), number of users (thousands), number of active users (users which have edited in the last month), and the percentage of edits
by unregistered users (known by their IP-addresses) to the all edits.
Two last columns consist of the assigned UTC offset to each WP and the Sleep Depth
respectively. The demographic data is taken from Wikipedia and supposed to give an impression to the reader. In the paper, there is not
any analysis based on this data.
\end{flushleft}
\label{tab:stat}
 \end{table}

\end{document}